\documentclass[a4paper,12pt]{article}
\usepackage[colorlinks,pdfpagelabels,pdfstartview=FitH,bookmarksopen=true,bookmarksnumbered=true,linkcolor=blue,plainpages=false,hypertexnames=false,citecolor=blue, urlcolor=blue]{hyperref}
\usepackage{amsfonts}
\usepackage{amsmath}
\usepackage{graphicx}
\usepackage{amssymb}
\usepackage{longtable}
\usepackage{authblk}
\usepackage{pdflscape}
\usepackage{rotating}
\usepackage[round]{natbib}
\usepackage{booktabs}
\usepackage{amsmath} % For equations and matrix environments
\usepackage{amssymb} % For mathematical symbols like \sim
\usepackage{amsthm} % If you use theorems, not strictly necessary here
\usepackage{bm} % For bold math symbols (e.g., bold matrices)
\usepackage[hmargin=2.5cm,vmargin=3.5cm]{geometry}
\usepackage{setspace}
\usepackage{chngcntr}
\usepackage{verbatim}
\usepackage{multirow}
\usepackage{subcaption}
\usepackage{chngcntr}
\usepackage[title]{appendix}
\usepackage[gen]{eurosym}
\usepackage{comment}
\usepackage[bottom]{footmisc}
\usepackage{gensymb}
\usepackage{textcomp,mathcomp}
\usepackage{wrapfig}
\usepackage{pdflscape}
\usepackage{rotating}
\usepackage{epstopdf}
\usepackage{graphicx}
\usepackage{caption}
\usepackage{subcaption}
\usepackage{longtable,tabularx,ltxtable,ragged2e}
\usepackage{tabularx, booktabs, makecell}
\begin{document}
%\pagenumbering{gobble}
\pagenumbering{arabic}
\author[1] {John Beirne}
\author[2] {Haroon Mumtaz}
\author[3] {Donghyun Park}
\author[4,5] {Gazi Salah Uddin}
\author[6] {Angeliki Theophilopoulou}
\affil[1] {Asian Development Bank}
\affil[2]{Queen Mary University of London}
\affil[3]{The SEACEN Centre}
\affil[4] {Linköping University}
\affil[5]{ Norwegian University of Life Sciences}
\affil[6]{Brunel University}
\title{Who benefits from increases in military spending? An empirical analysis.\footnote{Corresponding author: Haroon Mumtaz (\href{mailto:h.mumtaz@qmul.ac.uk}{h.mumtaz@qmul.ac.uk}). Do not circulate without authors' permission. The views expressed in this paper are solely the responsibility of the authors and should not be interpreted as reflecting the views of the  SEACEN Centre or the Asian Development Bank. }}

\date{This version: \today.}
\maketitle
\begin{abstract}
This paper investigates the heterogeneous effects of military spending news shocks on household income and wealth inequality for a large, panel of advanced and emerging economies. Confirming prior literature, we find that military spending news shocks lead to persistent increases in aggregate output and Total Factor Productivity. Our primary contribution is documenting contrasting distributional impacts. We find that expansionary military spending is associated with a mitigation of income inequality, as income gains are disproportionately larger at the left tail of the distribution, primarily driven by a rise in labour income and employment in industry. Conversely, the shock is found to increase wealth inequality, particularly in high-income countries, by raising the wealth share of the top decile ($\text{P}_{100}$) via effects on business asset holdings.


\end{abstract}

% \clearpage

% \linespread{1.2}\tableofcontents
% \linespread{1.2}\listoftables
% \linespread{1.2}\listoffigures

\clearpage
\pagenumbering{arabic} 
\setcounter{page}{2}

\section{\label{intro}Introduction}

Recent geopolitical events, such as the war in Ukraine, have brought the question of military spending to the forefront of the policy debate. According to the Stockholm International Peace Research Institute, \$2.7 trillion is directed at military expenditures in 2024 \cite{tian2024trends}.

Many commentators have argued that military preparedness is the key to prudent and effective defence policy. From an economic perspective, increases in military spending have been linked to long-run increases in output making this policy attractive to governments struggling with sluggish growth rates. An influential paper that makes this point is \cite{as2022}. Using a long spanning time-series of historical data for the US \cite{as2022} find that military spending news shocks have large and persistent effects on output because they shift the composition of public spending toward Research \& Development (R\&D). This surge in public R\&D spending boosts innovation and private investment in the medium term and subsequently increases productivity and GDP in longer horizons, causing the output multiplier to rise significantly above one in longer horizons. \cite{BenZeev2025Offensive} report similar results for a panel of European countries. They find multipliers above unity with the shock transmitted by R\&D spending and TFP growth. They also report that the shock reduces labour share pointing to potential adverse distributional effects. Both \cite{as2022} and \cite{BenZeev2025Offensive} build on a large literature examining the role of military and government spending, in general, and estimating the fiscal multiplier (see \cite{PappaDefense} and \cite{RameyZubairy}, \cite{Auerbach} for prominent examples).
\cite{sheremirov2022fiscal} report evidence on the effects of government spending shocks using variation in military spending on output from a large panel of 36 advanced and 93 developing countries and found that substantial heterogeneity is observed across countries, during the period 1988–2013.
\newline Although there is strong evidence that military spending boosts \textit{aggregate} output in the medium- and long-run, evidence on the distributional effects of the shock is limited. Our paper fills this gap in the literature and considers whether military spending news shocks have a heterogeneous effect on income and wealth across households. We carry out this investigation for a large panel of advanced economies and emerging markets. Our results suggest that expansionary military spending is associated with larger increases in income at the left tail of the distribution when compared to the right tail, and thus helps to mitigate income inequality. In contrast, there is evidence that in high-income countries, these shocks increase wealth disproportionally at the right tail of the distribution.
\newline The paper is related to the large literature that relates policy shocks to income and wealth inequality. For example \cite{Colciago_2019}, \cite{COIBION201770}, \cite{MUMTAZ2017410} and \cite{repec:eee:eecrev:v:130:y:2020:i:c:s0014292120302282} investigate the link between monetary policy and inequality. In a recent paper, \cite{furceri2022} examined the link between fiscal policy and inequality for a large panel of countries. They find that fiscal expansions are associated with lower inequality.\footnote{Older papers that examine the effects of government spending shocks on inequality include \cite{wolff2007distributional} for the US and \cite{ramos2008} for the UK. \cite{repec:pra:mprapa:123457} investigate the distributional effects of tax shocks in the UK.} Our paper adds to this literature in four ways. First, it focuses on the effect of military spending news shocks rather than government spending in general. Second, we investigate the effect of these shocks on the distribution of both income and wealth. Third, by using panel VARs with long lags, we examine the impact of these shocks at medium and long-horizons. Finally, our data set covers both developed and emerging markets, allowing us to investigate whether the distributional effect of military news shocks varies with country characteristics.
\newline The paper is organised as follows: Section \ref{sec:model} introduces the empirical approach, the data set is described in section \ref{subsec:data} and the main results are presented in section \ref{sec:results}.

\section{Empirical Model} \label{sec:model}

We estimate a Bayesian Panel VAR model:
\begin{equation}
    Z_{it}=\alpha_{i}+\tau_{t}+\sum_{l=1}^{L}B_{l}Z_{it-l}+A_{0}\varepsilon_{it}
\end{equation}
where $i=1,2,\dots M$ indexes the countries, $\alpha_i$ are fixed effects and $\tau_t$ denote time effects.  $Z_{t}$ are the endogenous variables. These include four aggregate variables for each country: (i) Real military spending per-capita, (ii) Real GDP per-capita, (iii) primary balance to GDP ratio and (iv) Labour productivity. To this basic set we add $10$ distributional variables: real average pre-tax income \textit{or} net wealth in the deciles of the respective distributions. These two $14$ variable VARs constitute the benchmark models.
As in \cite{as2022}, we set the lag length to $15$. The use of long lags is essential to ensure that the VAR impulse responses (IRFs) capture medium and long-run dynamics accurately (see \cite{baumeister_discussion_2025} and \cite{WestermarkDeGraeve:2025}).  We follow \cite{as2022} and \cite{RePEc:jae:japmet:v:25:y:2010:i:1:p:71-92} and use a Minnesota type prior on the VAR coefficients. \footnote{We use a standard value of $0.2$ for the tightness of the prior. As shown in the robustness checks, we obtain results that are similar to benchmark when using a looser prior.}
 \cite{RePEc:jae:japmet:v:25:y:2010:i:1:p:71-92} state the posterior distributions of the VAR parameters and describe a simple MCMC algorithm to draw from these distributions. Details of the algorithm can be found in the appendix to the paper.

\subsection{Identification}
We identify military spending news shocks using the strategy devised by \cite{PappaDefense}. \cite{PappaDefense} builds on \cite{Barsky-Sims-11} and assumes that military spending news shocks are those that explain the bulk of military spending movement over a $5$ year horizon while leaving the current value of this variable unchanged. The identification scheme is consistent with the view that military spending is contemporaneously unaffected by other variables and driven by two shocks: an unanticipated change in spending that has a contemporaneous affect and a news shock that moves spending after a lag. Given rational expectations, other variables may react immediately to the military spending news shock implying that this disturbance can be recovered in a multi-variate VAR setting.
More formally, the moving average representation links the endogenous variables $\mathbf{Z}_{it}$ to the vector of orthogonal, unit variance structural shocks $\mathbf{\varepsilon}_{it}$:
\begin{equation}
\label{eq:MA_rep}
\mathbf{Z}_{it} = \mathbf{C}(L)\mathbf{\varepsilon}_{it} = \sum_{j=0}^{\infty} \mathbf{C}_{j} \mathbf{\varepsilon}_{i, t-j}
\end{equation}
where $\mathbf{C}_{j}$ is the $j$-th matrix of common impulse responses. Let $g_{it}$ denote the variable for real military spending per capita (the $m$-th element of $\mathbf{Z}_{it}$). The news shock, $\varepsilon^N_{it}$ (the $k$-th element of $\mathbf{\varepsilon}_{it}$), is identified by two conditions. First, the shock must maximize the contribution to the forecast error variance of $g_{it}$ over a chosen horizon $H$. The optimization problem is given by:
\begin{equation}
\label{eq:MFEV_maximization}
\max_{\mathbf{A_k}} \frac{\sum_{j=0}^{H-1} (\mathbf{C}_j)_{m, k}^2}{\sum_{l=1}^{K} \sum_{j=0}^{H-1} (\mathbf{C}_j)_{m, l}^2}
\end{equation}
where $K$ is the total number of structural shocks (the dimension of $\mathbf{Z}_{it}$). The denominator of the fraction in equation \eqref{eq:MFEV_maximization} defines the variance of the $H$-step ahead forecast error for the variable $g_{it}$:
\begin{equation}
\label{eq:variance_g}
\text{Var}(g_{i, t+H} - E_t[g_{i, t+H}]) = \sum_{l=1}^{K} \sum_{j=0}^{H-1} (\mathbf{C}_j)_{m, l}^2
\end{equation}
Second, the news shock must have no contemporaneous effect on military spending. This requires the element of the impact matrix $\mathbf{C}_0 = \mathbf{A}$ corresponding to the effect of the news shock ($k$) on current military spending ($m$) to be zero:
\begin{equation}
\label{eq:contemp_restriction}
(\mathbf{C}_0)_{m, k} = 0
\end{equation}
This Maximum Forecast Error Variance (MFEV) method, subject to the orthogonality constraint in Equation \ref{eq:contemp_restriction}, essentially selects the linear combination of the current period's VAR innovations that best predicts the future path of military spending over the defined horizon $H$.

\section{Data} \label{subsec:data}
The data set covers the $83$ countries listed in Table \ref{tab:sample_periods_full}. The time-series is annual with a maximum span from 1960 to 2023. As shown in Table \ref{tab:sample_periods_full} the panel is unbalanced. However, for all the countries in the sample, the data is available at least for $20$ years.
We obtain data on nominal military spending, population, real GDP and the GDP deflator from World Banks' \href{https://databank.worldbank.org/source/world-development-indicators}{World Development Indicators}. Nominal military spending is deflated by the GDP deflator and both real spending and GDP are divided by population. We include logs of these variables in the VAR model. The primary balance to GDP ratio is obtained from the \href{https://www.imf.org/external/datamapper/pb@FPP/ITA} {IMF data base}. We obtain TFP from the \href{https://www.rug.nl/ggdc/productivity/pwt/?lang=en}{Penn World tables}.
Data on the distribution of income and wealth is taken from the \href{https://wid.world/}{World Inequality Database (WID)}. Our benchmark measure of income is \textit{pre-tax National Income (code:PTINC)}. This is defined as income that includes social insurance benefits net of corresponding contributions but excludes income tax, social assistance benefits and other forms of redistribution. We obtain data on this variable in $10$ percentile groups for each year: Average income below the $10^{th}$ percentile ($P_{10}$), average income between the $10^{th}$ sand $20^{th}$ percentile ($P_{20}$) and so on until the final group $P_{100}$ that denotes average income above the $90^{th}$ percentile. Average income in each group is deflated using the GDP deflator and we include logs of these variables in the VAR model.
We also obtain the wealth distribution for each country from WID. The measure of wealth is \textit{Net Household Wealth (Code:HWEAL)}. We include the share of total wealth in each of the percentile groups $P_{10},P_{20},\dots,P_{100}$ in the VAR model.
We also use additional aggregate variables, where we discuss the transmission mechanisms of the military spending shocks. Details of the sources for these additional series are presented in the appendix.

\section{Results} \label{sec:results}
\begin{figure}
    \centering
    \includegraphics[width=1\linewidth]{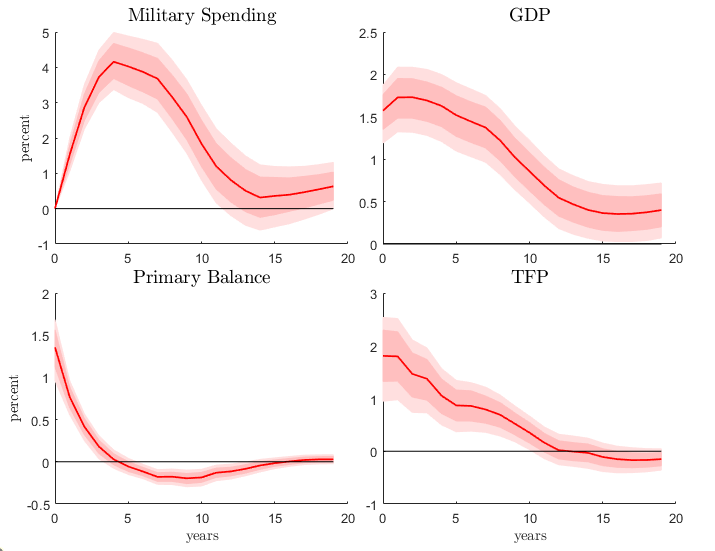}
    \caption{Impulse response of Macroeconomic variables to a military spending news shock. The dark (light) shaded area is the 68\% (90 \%) error band. }
    \label{fig:irf_macro}
\end{figure}

Figure \ref{fig:irf_macro} presents the response of the aggregate macroeconomic variables to one standard deviation military spending news shock from our benchmark $14$ variable model that includes average income in groups $P_{10}$ to $P_{100}$.\footnote{The responses of these variables from the model that includes the distribution of wealth are very similar.} Military spending rises by about $4\%$ five years after the shock. Real GDP rises on impact and the increase persists in the long-run. The shock induces a persistent increase in TFP providing support for the supply-side transmission of the shock documented in \cite{BenZeev2025Offensive} and \cite{as2022}.

\begin{landscape}
\begin{figure}
    \centering
    \includegraphics[width=0.9\linewidth]{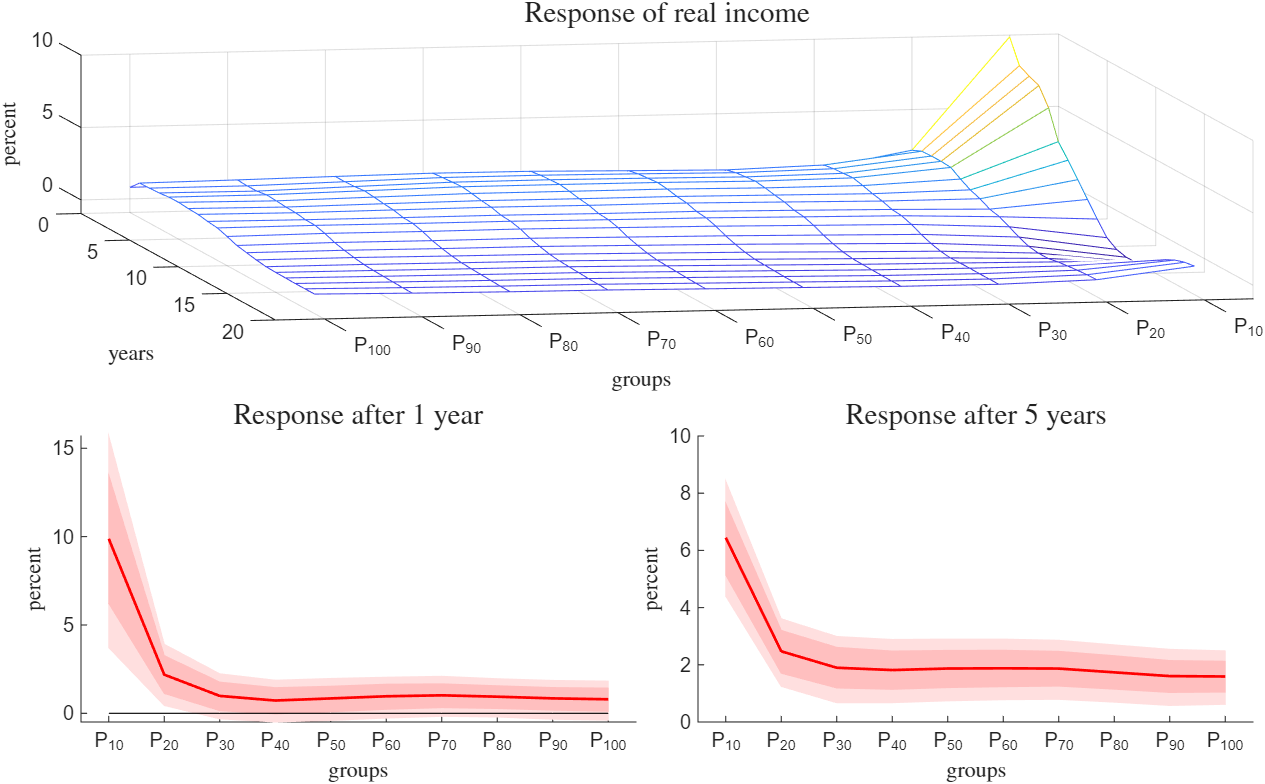}
    \caption{Impulse response of the income distribution to a military spending news shock. The dark (light) shaded area is the 68\% (90 \%) error band. }
    \label{fig:irf_income}
\end{figure}
\end{landscape}

\begin{landscape}
\begin{figure}
    \centering
    \includegraphics[width=0.9\linewidth]{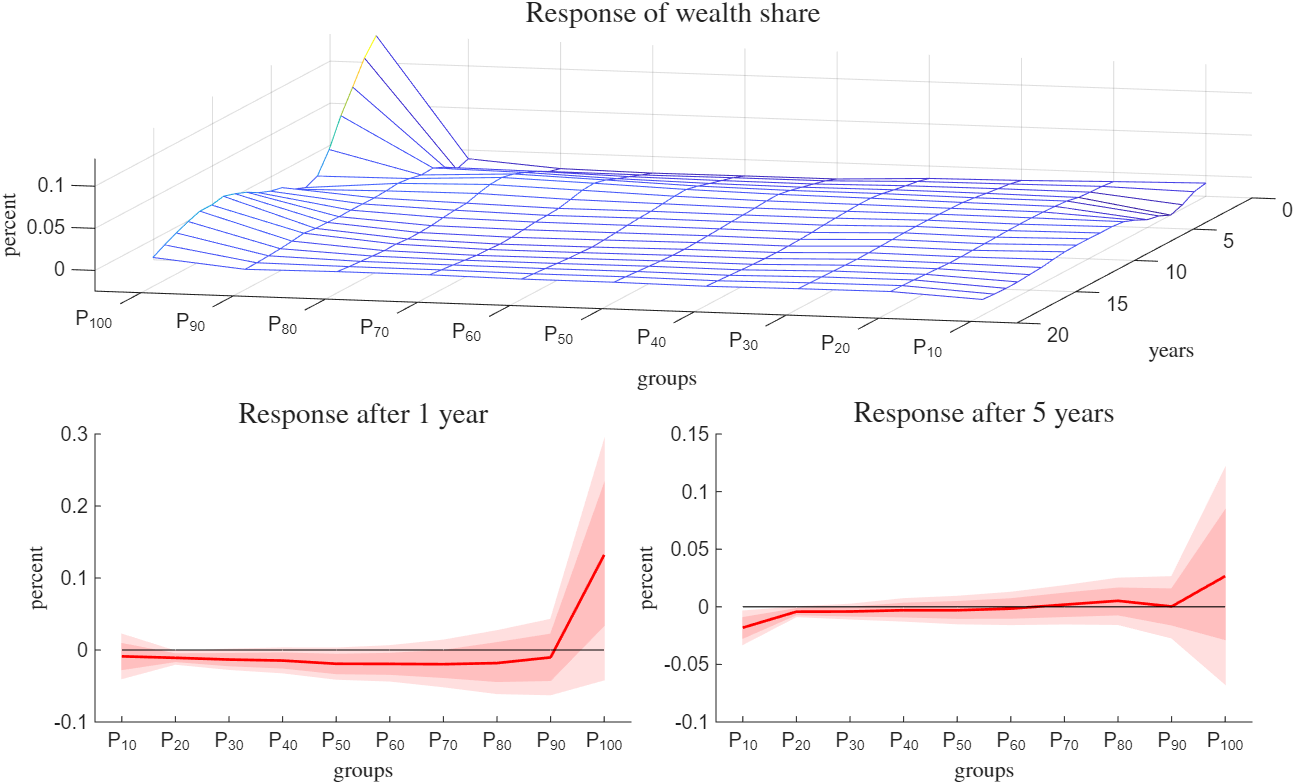}
    \caption{Impulse response of the wealth distribution to a military spending news shock. The dark (light) shaded area is the 68\% (90 \%) error band. }
    \label{fig:irf_wealth}
\end{figure}
\end{landscape}
Figure \ref{fig:irf_income} displays the response of average real income in each decile group. The top panel shows the median response, while the bottom panels depict the response at different horizons together with the error bands. The news shock has a positive impact on income across the distribution. However, it is clear that the impact is largest at the left tail of the income distribution. For example, at the $5$ year horizon, average income in the first decile rises by about $6\%$. For group $P_{20}$, this increase is about $2.5\%$ while the effect for $P_{100}$ is $1.6\%$

Next we re-estimate the benchmark model replacing average income with wealth shares in each decile group. Figure \ref{fig:irf_wealth} presents the response of wealth shares in each group to the military news shock. The response of the distribution of wealth stands in sharp contrast to the estimated response of income in the decile groups. The response of the wealth share towards the left tail of the distribution is minimal with some evidence of a decline for groups $P_{10}$ and $P_{20}$ at long horizons. The wealth share of $P_{30}$ to $P_{90}$ is largely unaffected by the shock. However, for group $P_{100}$, the shock induces an increase in the wealth share of about $0.1\%$ at the one year horizon with $0$ not included in the $68\%$ credible intervals. We show in the analysis below that this heterogenous response of the top $10\%$ wealth share is especially prominent and precisely estimated for high-income countries.
\begin{figure}
    \centering
    \includegraphics[width=0.8\linewidth]{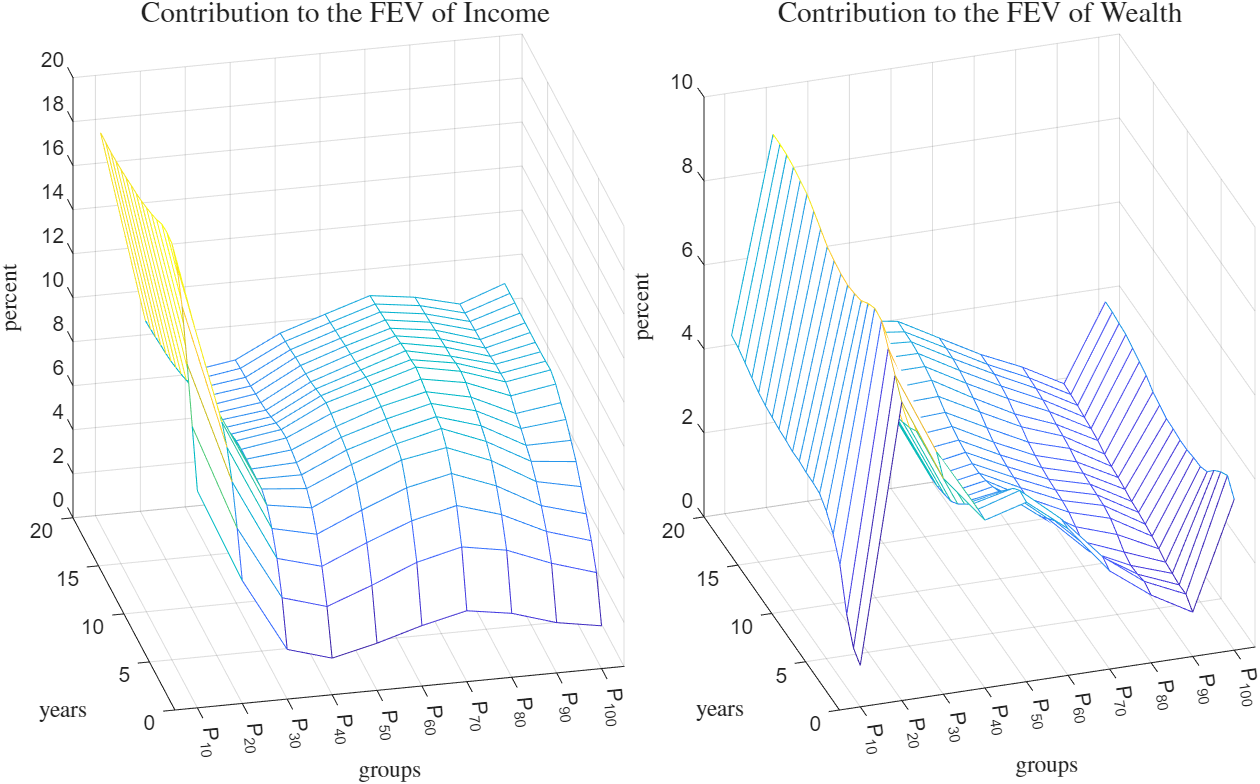}
    \caption{Contribution of the military spending news shock to the forecast error variance of income and wealth in decile groups.  }
    \label{fig:fev}
\end{figure}

Figure \ref{fig:fev} presents the contribution of the military spending news shock to the forecast error variance (FEV) of the distributional income and wealth variables. The left panel of Figure \ref{fig:fev} shows that the military spending news shock makes the largest contribution to the FEV of income in the first two decile groups which is estimated to be above $10\%$. The importance of this shock for income in the remaining groups is more modest. The right panel of the figure shows that news shock makes a relatively larger contribution to the FEV of net wealth in groups $P_{20}$ and $P_{100}$. While the magnitude of the contributions are smaller than that for income, these results highlight that the shock hold some importance for both tails of the wealth distribution.
\subsection{Heterogeneity}
As our data set spans a large set of countries and regions, we consider if the distributional responses presented above vary with key country characteristics.
\begin{figure}
    \centering
    \includegraphics[width=0.9\linewidth]{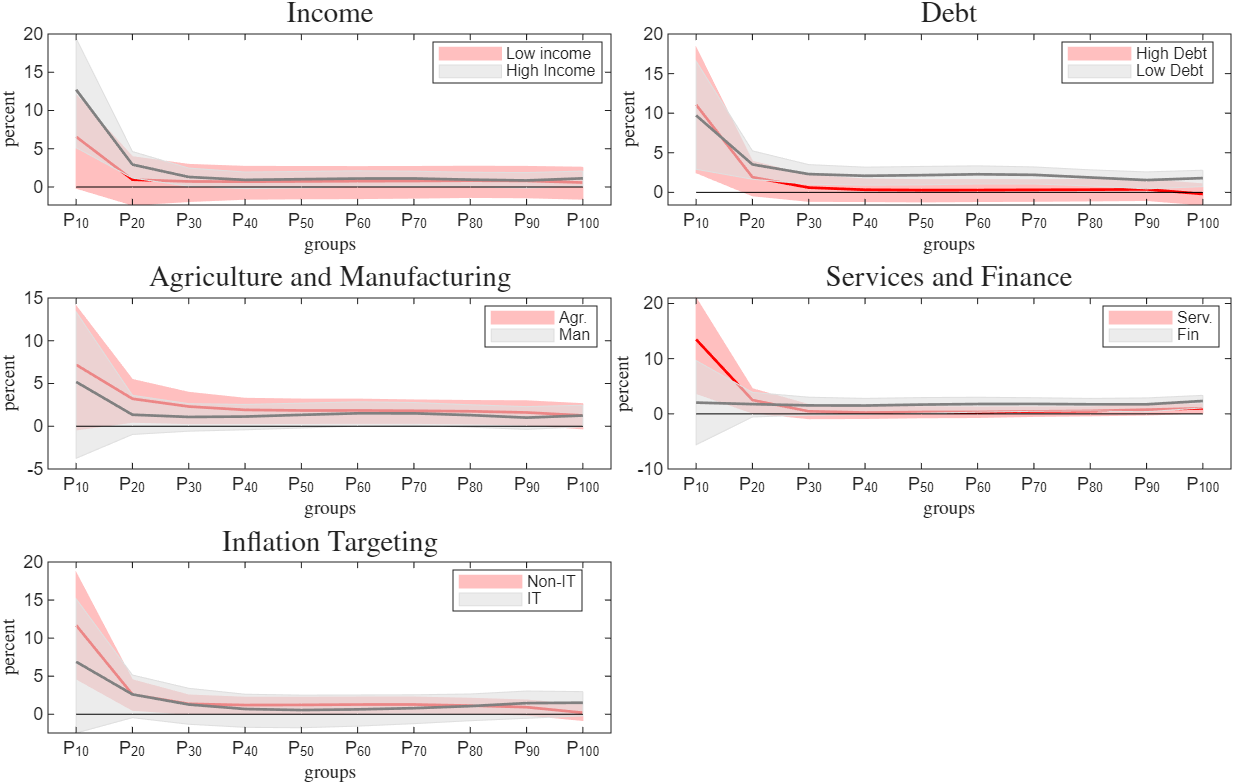}
    \caption{The response of income deciles for different sets of countries. The shaded area shows the 90\% error bands.}
    \label{fig:het_income}
\end{figure}
The top left panel of Figure \ref{fig:het_income} shows the IRFs from the benchmark VAR model, estimated separately for countries classified by the World Bank as either high income or higher middle income, and those those classified as low or lower-middle income. While the pattern of the IRFs is similar across groups, there is some evidence that the impact of the military news shock on the left tail of the distribution is more pronounced in high income countries. The top right panel shows that the level of general government debt to GDP ratio has a limited impact on the response of $P_{10}$ to the shock. However, income for $P_{20}$ to $P_{100}$ rises more in low debt countries (i.e. countries with lower than median debt to GDP ratios in 2019) suggesting a smaller impact on income inequality for this group. The middle panel of the figure classifies countries on the basis of the composition of industrial sectors. For example, agricultural countries are those that have a larger than median share in this sector in 2019. The results from this exercise indicate that the income of households in the left tail tends to experience a positive boost in countries with large agricultural, manufacturing, and services sectors, whereas the benchmark results for income are least visible in countries with a large financial sector. Finally, the bottom panel of the figure shows that the median effect on the left tail of the income distribution is smaller in countries with inflation targeting regimes, possibly indicating that the response of monetary authorities to the fiscal shock may also play a role in determining the outcome for the left tail income group.

\begin{figure}
    \centering
    \includegraphics[width=0.9\linewidth]{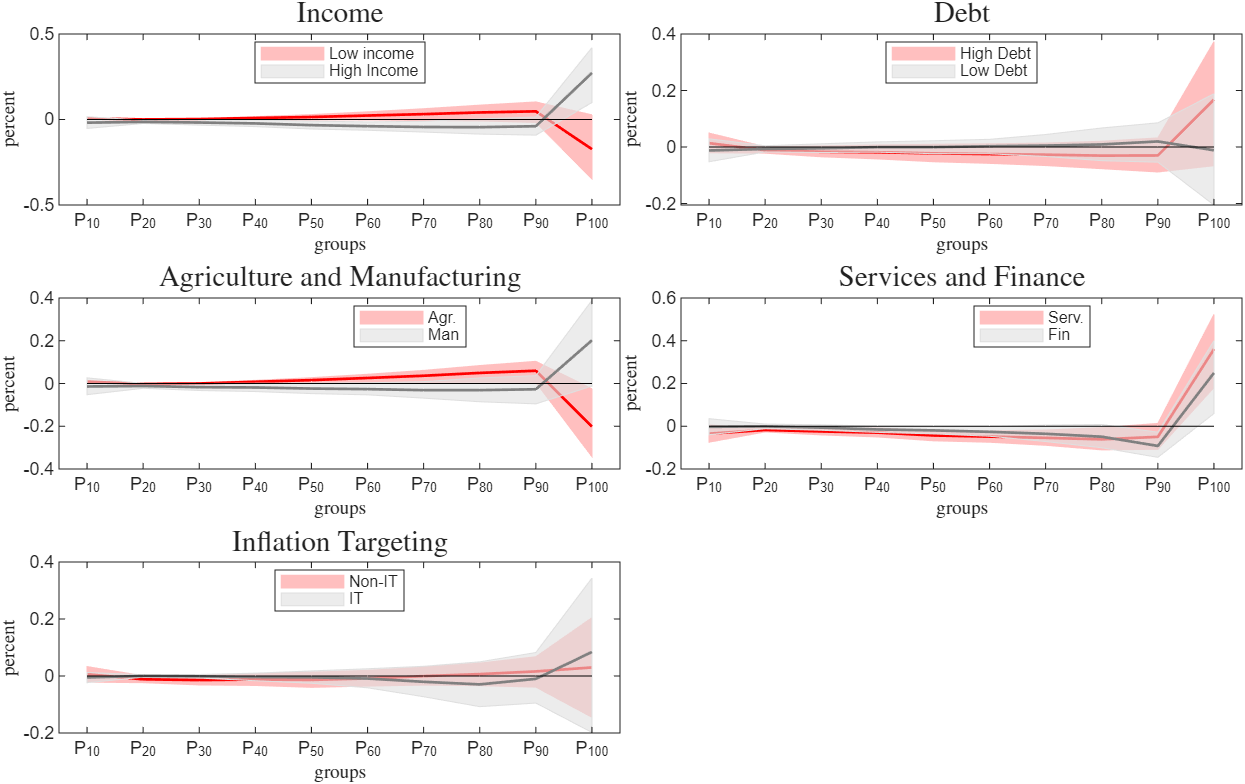}
    \caption{The response of wealth shares for different sets of countries. The shaded area shows the 90\% error bands.}
    \label{fig:het_wealth}
\end{figure}

Figure \ref{fig:het_wealth} shows the results for the benchmark panel VAR for wealth estimated on different groups of countries. It is clear from the figure that the cross-country heterogeneity is more pronounced for the response of the wealth distribution. The top left panel of the figure shows that the impact of the military news shock on the wealth distribution is substantially different across high and low income countries. In high income economies, the shock is associated with an increase in the share of wealth for the top decile, and this rise is statistically different from zero when considering the 90\% error bands . In contrast, in low income countries, the median response of wealth in groups $P_{60}$ to $P_{90}$ is positive and that of $P_{100}$ is negative. Note, however, that the error bands include zero across the groups, indicating the wealth distribution is largely unaffected in low income countries. This result is also reflected in the sectoral split. The shock has a positive impact on the top wealth decile in countries with larger manufacturing, service and financial sectors, while the response of the wealth distribution in agricultural countries shows a pattern similar to that estimated for low income economies. The splits by the level of debt and the inflation targeting regime, however, are largely uninformative due to the wide error bands.

\subsection{Transmission Channels}
\subsubsection{Income}
\begin{figure}
    \centering
    \includegraphics[width=0.8\linewidth]{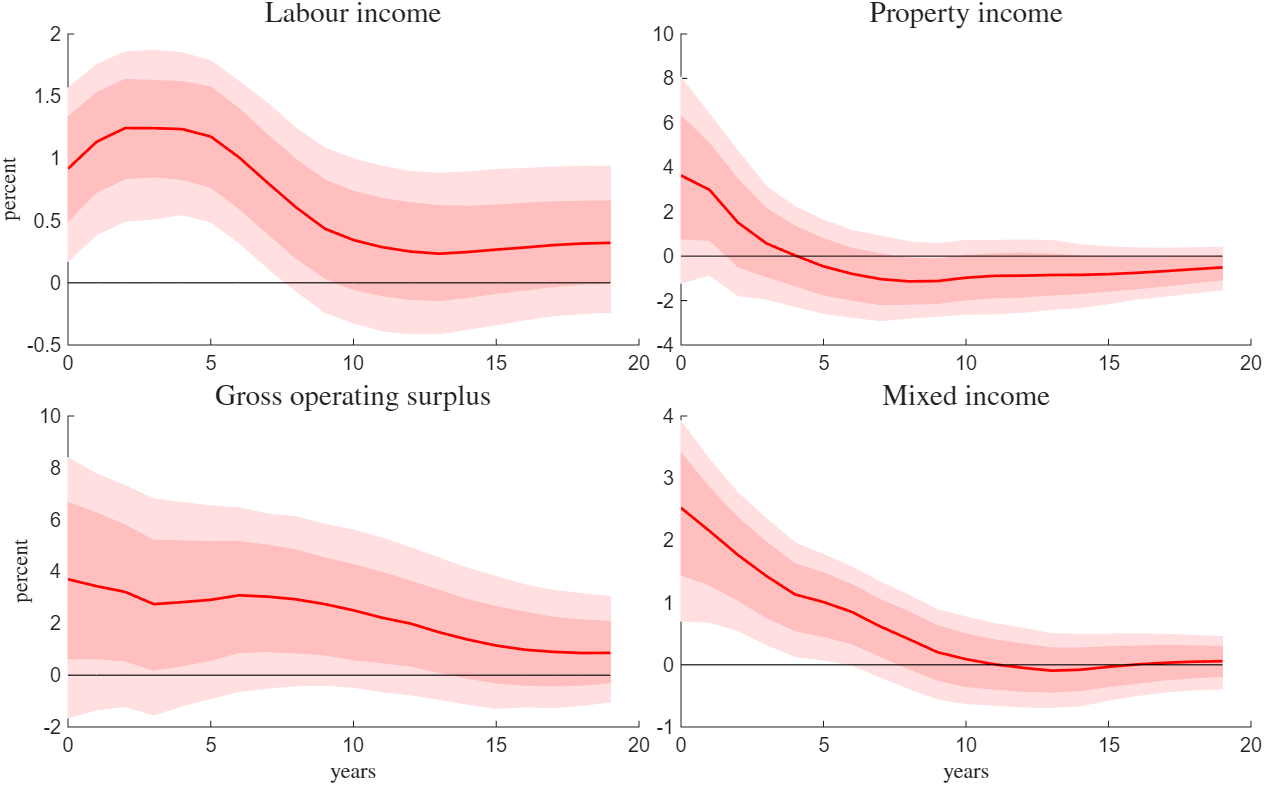}
    \caption{Response of Labour income components to the military spending news shock }
    \label{fig:irf_income_comp}
\end{figure}

To investigate the transmission of the military news shock to income, we examine the response of the components of income of households and  non-profit institutions serving households (NPISH) and that of key labour market variables. We first estimate a panel VAR model that includes the benchmark aggregate variables (military spending, GDP, primary balance, and TFP) and add four components of aggregate real income obtained from WID: (1) labour income, (2) net property income, (3) net operating surplus (rental income, including imputed rents) and (4) net mixed income (income from self employment). 
Figure \ref{fig:irf_income_comp}  shows that the shock has a statistically non-zero impact on labour income and income from self-employment, leading to a persistent increase in both variables. In contrast, the shock does not seem to affect components of income that are likely to be more prevalent at the right tail of the income distribution (property income and rents). 
Next, We consider the response of labour market quantities. We estimate a version of the panel VAR where the basic $4$ endogenous variables are augmented with unemployment as a share of total labour force and the shares of employment in industry, agriculture and services, respectively. All of these variables are obtained from the World Bank data base.

\begin{figure}
    \centering
    \includegraphics[width=0.8\linewidth]{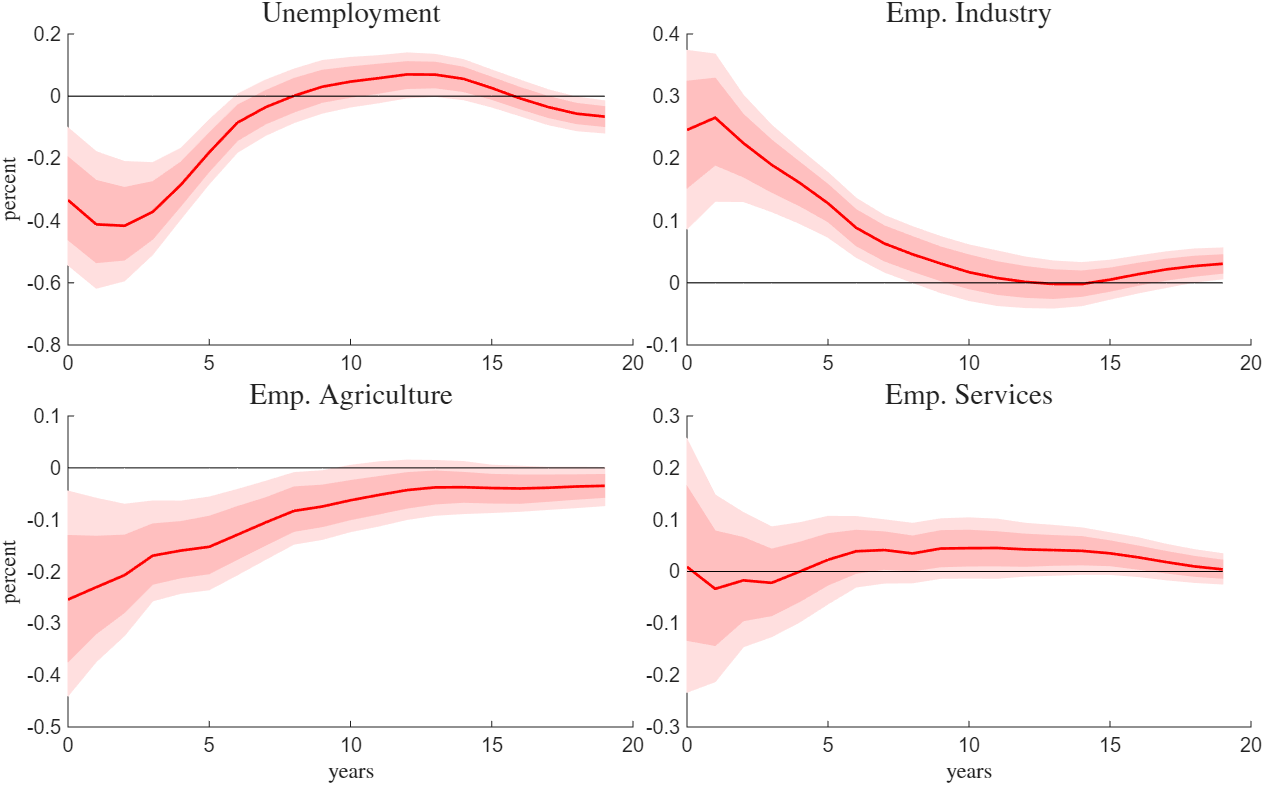}
    \caption{Response of Labour market variables to the military spending news shock }
    \label{fig:irf_labour}
\end{figure}

The top left panel of Figure \ref{fig:irf_labour} shows that the shock is associated with a large fall in Unemployment indicating a boost to demand for labour. The remaining responses show that employment rises in industry while declining in the agricultural sector and remaining broadly stable in services. \footnote{ The industry sector consists of mining and quarrying, manufacturing, construction, and public utilities (electricity, gas, and water).} This is consistent with a shift of workers towards sectors such as manufacturing and construction that are likely to be boosted by the military news shock. As blue collar and manual workers are concentrated in these sectors, this increased demand may explain the rise in labour income estimated for groups $P_{10}$ and $P_{20}$.
\subsubsection{Wealth}
The pattern of the net wealth response across the distribution suggests that the military news shock affects components of wealth more prevalent towards the right tail of the distribution. To investigate this further we estimate the panel VAR model augmenting the basic $4$ endogenous variables with the components of net wealth of households and NPISH obtained from WID. These include components of non-financial wealth and financial wealth. The former constitutes housing assets and business and non-financial assets. Financial wealth contains (1) currency, deposits, bonds, and loans, (2) equities and fund shares and (3) pension funds and life insurance.\footnote{Financial wealth also includes off-shore wealth. However, data on this variable is unavailable for the majority of countries.}. These $5$ wealth components are expressed as percentages of total net wealth and included in the VAR. Given the strong heterogeneity in the response of wealth across countries with different levels of income, we estimate this panel VAR separately for high and low income countries.
\begin{figure}
    \centering
    \includegraphics[width=0.8\linewidth]{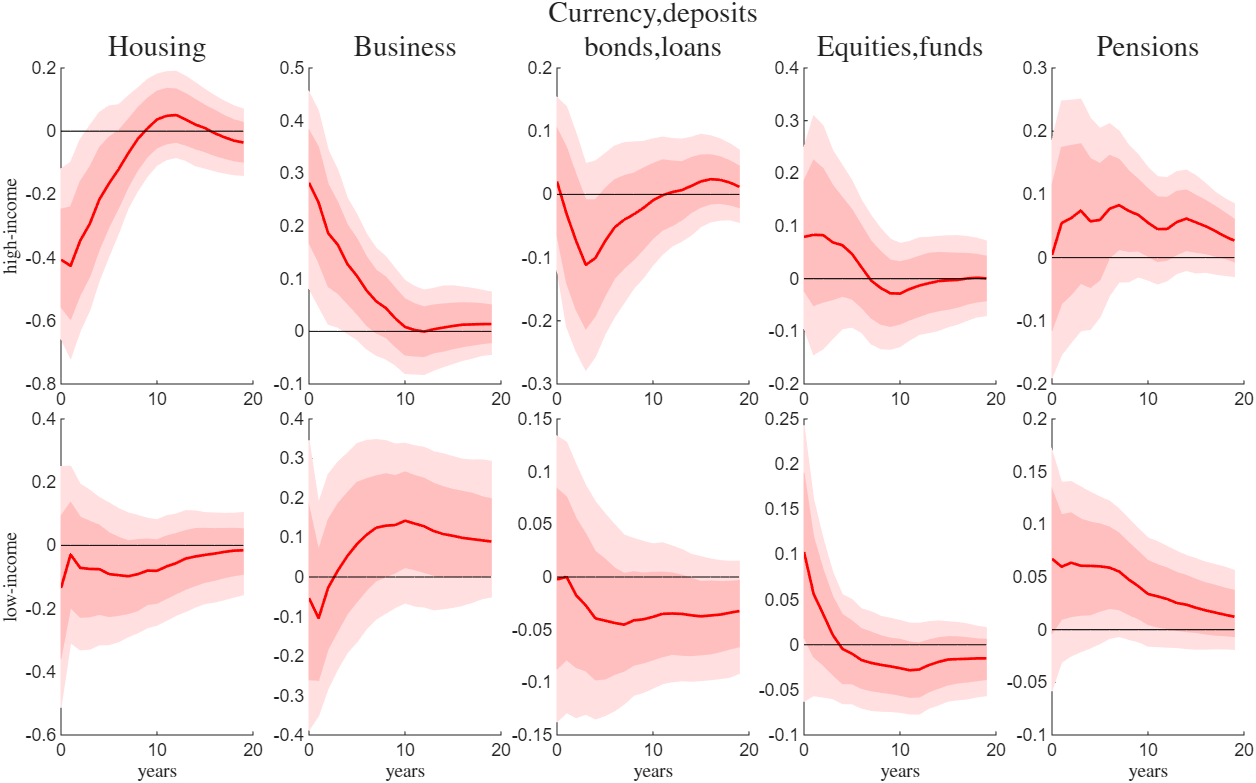}
    \caption{Response of wealth components to the military spending news shock }
    \label{fig:irf_wealth_comp}
\end{figure}
The bottom panel of Figure \ref{fig:irf_wealth_comp} shows that the response of wealth components in low-income countries is imprecisely estimated. In high-income countries, however, the shock is clearly associated with an increase in the share of wealth associated with business assets and a decline in the share of housing wealth. These results suggest that the increase in wealth at the right tail of the distribution may be driven via the effect of the military news shock on non-financial business assets.
\subsection{Robustness}
\begin{figure}
    \centering
    \includegraphics[width=1.1\linewidth]{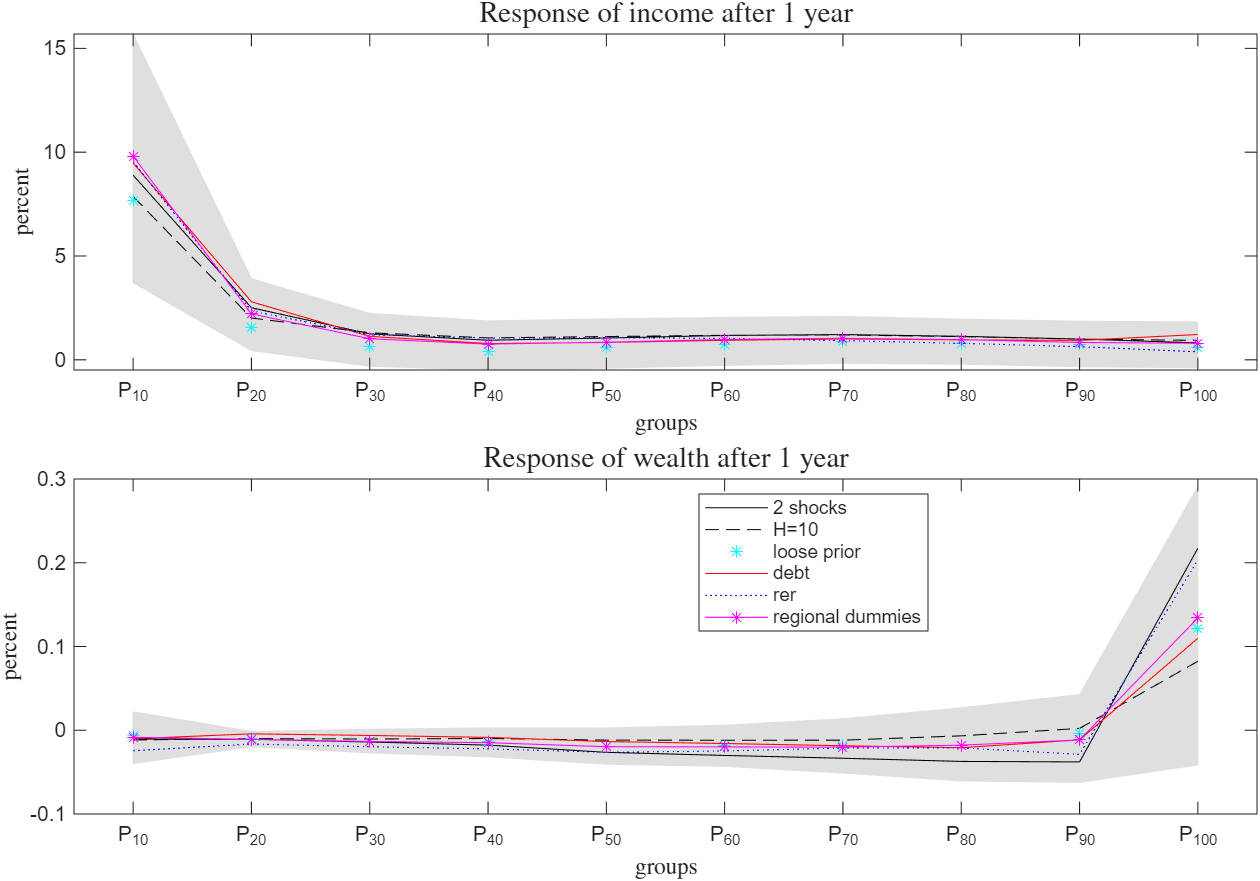}
    \caption{Robustness analysis. The grey shaded area represents the 90\% error bands obtained using the benchmark models. The median estimates from alternative models are shown as lines and symbols. }
    \label{fig:irf_robust}
\end{figure}
In this section we present results from a comprehensive robustness analysis. The results are presented in Figure \ref{fig:irf_robust}.
\subsubsection{Identification} \label{sec:identification}
One potential concern with the identification is that it may be picking up other shocks that lead to current and future changes in GDP. A candidate for such a confounding shock is TFP news. In order to investigate this further we estimate a version of the VAR where we identify both TFP news and military spending news shocks and restrict the two disturbances to be orthogonal. We identify the TFP news shock by imposing the condition that this shock explains the bulk of the variation in TFP over a horizon of $5$ years while leaving current TFP unchanged. The military news shock is identified using the benchmark assumptions on the response of military spending discussed above, albeit with the additional requirement that this shock is orthogonal to TFP news. As shown by the solid black lines in Figure \ref{fig:irf_robust}, they key results survive in this more elaborate setting. 
We also consider a version of benchmark model where the restrictions on the FEV of military spending are imposed at a horizon of $10$ years. This change has minimal effects on the key results (dotted black lines in Figure \ref{fig:irf_robust}).
\subsubsection{Specification}
As discussed above, we employ a Minnesota prior for the panel BVAR. We consider the sensitivity of the results to prior tightness by re-estimating the benchmark models setting the tightness parameter to $1$ instead on the benchmark value of $0.2$. The blue dots in \ref{fig:irf_robust}) show that the pattern of the distributional responses are preserved in this alternative setting. Adding regional dummies to the models does not change the results. Finally, adding the debt to GDP ratio or the real exchange rate does not materially alter the main conclusions reached above.
\section{Conclusion}
Our analysis reveals a contradictory distributional impact of military spending news shocks. While these shocks act as a temporary equaliser for income by boosting employment in blue-collar sectors, they simultaneously exacerbate wealth inequality through capital gains favouring the top echelon, particularly in advanced economies. This finding highlights a crucial trade-off for policymakers: relying on military expenditure for aggregate demand stimulus may unintentionally worsen the long-term structural problem of wealth concentration. Future research should focus on modelling the explicit general equilibrium mechanism that links military spending to non-financial business assets to fully understand this wealth channel.

\newpage

\bibliographystyle{ecta}
\bibliography{references/references, references/references1,references/references1zz,references/LR_biblio,references/LR_biblio1,references/LR_biblio2,references/LR_biblio25,references/LR_biblio3,references/LR_biblio4,references/LR_bibliox,references/referenceszz,references/biblio,references/bibliography}

\newpage
\setcounter{footnote}{0}
\renewcommand{\thefootnote}{\arabic{footnote}}
\section{Appendix}
\subsection{Estimation of the panel BVAR model}
The panel VAR model is defined as: 
\begin{equation*}
Z_{it}=\alpha_{i}+\tau_{t}+\sum_{l=1}^{L}B_{l}Z_{it-l}+v_{it}
\end{equation*}%
where $\mathrm{var}\left( v_{it}\right) =\Omega $, $i=1,2,\dots,M$ indexes the countries
in our panel, $t=1,2,\dots,T$ denotes the time-periods. Let $%
\bm{b}=\mathrm{vec}\left( 
\begin{array}{c}
B_{1} \\ 
\vdots \\ 
B_{P} \\ 
\bm{c}%
\end{array}%
\right) $ where the vector $\underset{EX\times 1}{\underbrace{\bm{c}}}=\mathrm{vec}\left( 
\begin{array}{c}
\alpha _{i} \\ 
\tau _{t}%
\end{array}%
\right) $ collects the exogenous regressors (i.e. the time and country fixed effects).

\subsection{Priors}
\label{subsec:priors}

We follow \cite{RePEc:jae:japmet:v:25:y:2010:i:1:p:71-92} and use a Natural Conjugate prior
implemented via dummy observations. The priors are implemented by the dummy
observations $y_{D}$ and $x_{D}$ that are defined as:

\begin{equation}
y_{D}=\left[ 
\begin{array}{c}
\frac{\mathrm{diag}\left( \gamma _{1}s_{1},\dots,\gamma _{n}s_{n}\right) }{\kappa } \\ 
0_{N\times \left( P-1\right) \times N} \\ 
\mathrm{diag}\left( s_{1},\dots,s_{n}\right) \\ 
\hdotsfor{1} \\ 
0_{EX\times N}%
\end{array}%
\right] ,\ \ \ \ \ \ x_{D}=\left[ 
\begin{array}{c}
\frac{J_{P}\otimes \mathrm{diag}\left( s_{1},\dots,s_{n}\right) }{\kappa }\text{ }%
0_{NP\times EX} \\ 
\hdotsfor{1} \\ 
\text{ }0_{N\times (NP)+EX} \\ 
\hdotsfor{1} \\ 
0_{EX\times NP}\text{ \ \ \ \ \ \ \ \ }I_{EX}\times 1/c%
\end{array}%
\right]
\label{eq:yD_xD}
\end{equation}%
where $J_{P}=\mathrm{diag}(1,2,\dots,P)$, $\gamma _{1}$ to $\gamma _{n}$ denote the
prior mean for the parameters on the first lag obtained by estimating
individual AR(1) regressions, $s_{1}$ to $s_{n}$ is an estimate of the
variance of the endogenous variables obtained individual AR(1) regressions, $%
\kappa $ measures the tightness of the prior on the autoregressive VAR
coefficients, and $c$ is the tightness of the prior on the remaining
regressors. We set $\kappa =1$ and $c=1000$. We also implement priors on the
sum of coefficients (see \cite{RePEc:jae:japmet:v:25:y:2010:i:1:p:71-92}). The dummy observations
for this prior are defined as:%
\begin{equation}
\tilde{y}_{D}=\frac{\mathrm{diag}\left( \gamma _{1}\mu _{1},\dots,\gamma _{n}\mu
_{n}\right) }{\tau },\quad \tilde{x}_{D}=\left( 
\begin{array}{cc}
\left( 1_{1\times P}\right) \otimes \frac{\mathrm{diag}\left( \gamma _{1}\mu
_{1},\dots,\gamma _{n}\mu _{n}\right) }{\tau } & 0_{N\times EX}%
\end{array}%
\right)
\label{eq:ytildeD_xtildeD}
\end{equation}%
where $\mu _{i}$ is the sample average of the $i^{th}$ variable. As in \cite%
{RePEc:jae:japmet:v:25:y:2010:i:1:p:71-92} we set $\tau =10\kappa $. In the benchmark model $\kappa=0.2$ and $c=1000$

\subsection{Posterior and MCMC algorithm}
\label{subsec:posterior}

\cite{RePEc:jae:japmet:v:25:y:2010:i:1:p:71-92} show that posterior distribution can be written
as:%
\begin{equation}
g\left( \Omega |Y\right) \sim iW\left( \bar{\Omega},NT+2+NT-K\right)
\label{pos1}
\end{equation}

\begin{equation}
g\left( \bm{b}|\Omega ,Y\right) \sim N\left( \bar{\bm{b}},\Omega \otimes \left(
X_{\ast }^{\prime }X_{\ast }\right) ^{-1}\right)   \label{pos2}
\end{equation}%
where $iW$ denotes the inverse Wishart distribution, $NT$ is the total
number of observations, $K$ denotes the number of regressors in each
equation of the VAR model. Note that $Y_{\ast }=\left( 
\begin{array}{c}
Y \\ 
y_{D} \\ 
\tilde{y}_{D}%
\end{array}%
\right) $ and $X_{\ast }=\left( 
\begin{array}{c}
X \\ 
x_{D} \\ 
\tilde{x}_{D}%
\end{array}%
\right) $ and 
\begin{align*}
\tilde{B} &=\left( X_{\ast }^{\prime }X_{\ast }\right) ^{-1}\left( X_{\ast}^{\prime }Y_{\ast }\right)  \\
\bar{\bm{b}} &=\mathrm{vec}\left( \tilde{B}\right)  \\
\bar{\Omega} &=\left( Y_{\ast }-X_{\ast }\tilde{B}\right) ^{\prime }\left(
Y_{\ast }-X_{\ast }\tilde{B}\right) 
\end{align*}%
$Y$ and $X$ denote the matrices of dependent variables and regressors in the VAR equations, respectively. Posterior draws can be easily generated by drawing $\Omega $ from the
marginal distribution in \eqref{pos1} and then $\bm{b}$ from the conditional
distribution in equation \eqref{pos2}. We set the number of draws to 10,000
with a burn-in of 8,000. For each retained draw, we calculate the contemporaneous impact matrix via the method described in section \ref{sec:identification} and estimate the impulse response.

\subsection{Data}

\begin{table}[htbp]
    \centering
    \caption{Additional Data Sources}
    \label{tab:additional_data_sources}
    \begin{tabular}{|l|l|l|}
        \hline
        \textbf{Variable} & \textbf{Source} & \textbf{Code} \\
        \hline
        Debt to GDP & IMF & GC.DOD.TOTL.GD.ZS \\
        Real exchange rate & World Bank & PX.REX.REER \\
        Unemployment & World Bank & SL.UEM.TOTL.NE.ZS \\
        Employment industry & World Bank & SL.IND.EMPL.ZS \\
        Employment Agric. & World Bank & SL.AGR.EMPL.ZS \\
        Employment Serv. & World Bank & SL.SRV.EMPL.ZS \\
        \hline
        Net Wealth & WID & pweal \\
        Housing Assets & WID & pwhou \\
        Business and Non-Financial assets & WID & pwbus \\
        Currency, Deposits, Bonds and Loans & WID & pwfiw \\
        Equities and Funds & WID & pweqi \\
        Pension funds and life insurance & WID & pwpen \\
        Net Primary Income & WID & prihn \\
        Compensation of employees & WID & comhn \\
        Net Property income & WID & prphn \\
        Gross operating surplus & WID & gsrhn \\
        Mixed Income & WID & gmxhn \\
        \hline
    \end{tabular}
\end{table}

\begin{longtable}{lcc}
\caption{Full Sample Availability by Country (Total $N=83$)}
\label{tab:sample_periods_full}
\\
\toprule
\textbf{Country} & \textbf{Start Year} & \textbf{End Year} \\
\midrule
\endfirsthead

\multicolumn{3}{c}%
{{\bfseries \tablename\ \thetable{} -- continued from previous page}} \\
\toprule
\textbf{Country} & \textbf{Start Year} & \textbf{End Year} \\
\midrule
\endhead

\midrule
\multicolumn{3}{r}{{Continued on next page}} \\
\endfoot

\bottomrule
\endlastfoot

% --- FULL LIST OF 83 COUNTRIES EXTRACTED FROM 5569 ROWS ---
ALB & 1990 & 2023 \\
ARG & 1970 & 2023 \\
AUS & 1960 & 2013 \\
AUT & 1960 & 2023 \\
BEL & 1960 & 2023 \\
BGR & 1990 & 2023 \\
BHS & 1970 & 2019 \\
BLR & 1990 & 2023 \\
BRA & 1960 & 2023 \\
CAN & 1960 & 2023 \\
CHE & 1960 & 2023 \\
CHL & 1965 & 2023 \\
CHN & 1970 & 2023 \\
COL & 1965 & 2023 \\
CRI & 1970 & 2023 \\
CZE & 1993 & 2023 \\
DEU & 1991 & 2023 \\
DNK & 1960 & 2023 \\
DOM & 1970 & 2023 \\
ECU & 1970 & 2023 \\
EGY & 1970 & 2023 \\
ESP & 1965 & 2023 \\
FIN & 1960 & 2023 \\
FRA & 1960 & 2023 \\
GBR & 1960 & 2023 \\
GEO & 1995 & 2023 \\
GRC & 1965 & 2023 \\
GTM & 1970 & 2023 \\
HND & 1970 & 2023 \\
HRV & 1990 & 2023 \\
HUN & 1990 & 2023 \\
IDN & 1960 & 2023 \\
IND & 1975 & 2023 \\
IRL & 1960 & 2023 \\
ISR & 1965 & 2023 \\
ITA & 1960 & 2023 \\
JAM & 1970 & 2019 \\
JPN & 1960 & 2023 \\
KEN & 1965 & 2023 \\
KOR & 1960 & 2023 \\
KWT & 1970 & 2023 \\
LKA & 1960 & 2023 \\
LTU & 1990 & 2023 \\
LVA & 1990 & 2023 \\
MEX & 1980 & 2023 \\
MKD & 1995 & 2023 \\
MLT & 1970 & 2023 \\
MNE & 2006 & 2023 \\
MYS & 1960 & 2023 \\
NGA & 1960 & 2015 \\
NIC & 1970 & 2023 \\
NLD & 1960 & 2023 \\
NOR & 1960 & 2023 \\
NZL & 1960 & 2023 \\
OMN & 1970 & 2023 \\
PAN & 1970 & 2023 \\
PER & 1970 & 2023 \\
PHL & 1970 & 2023 \\
POL & 1990 & 2023 \\
PRT & 1975 & 2023 \\
PRY & 1970 & 2023 \\
QAT & 1970 & 2023 \\
ROM & 1990 & 2023 \\
RUS & 1992 & 2023 \\
SAU & 1980 & 2023 \\
SRB & 1995 & 2023 \\
SVK & 1993 & 2023 \\
SVN & 1995 & 2023 \\
SWE & 1960 & 2023 \\
THA & 1970 & 2023 \\
TTO & 1970 & 2023 \\
TUN & 1985 & 2023 \\
TUR & 1960 & 2023 \\
UKR & 1960 & 2013 \\
USA & 1960 & 2023 \\
URY & 1970 & 2023 \\
VEN & 1970 & 2017 \\
VNM & 1995 & 2023 \\
ZAF & 1965 & 2023 \\
ZMB & 1965 & 2023 \\
ZWE & 1970 & 2019 \\
\end{longtable}

\end{document}